\documentclass[conference]{IEEEtran}
\usepackage{graphicx}
\usepackage{xcolor}
\usepackage{amsfonts,amssymb}
\usepackage{subcaption}
\usepackage[linesnumbered,ruled,vlined]{algorithm2e}
\ifCLASSINFOpdf
\else
\fi
\usepackage{comment}
\usepackage{fancyhdr}
\usepackage{lastpage}
\usepackage{tabularx}
\usepackage{amsmath}
\usepackage{array,xcolor,colortbl}

\usepackage{multirow}
\usepackage{booktabs}
\DeclareUnicodeCharacter{2212}{-}
\begin{document}

\title{Relation-aware based Siamese Denoising Autoencoder for Malware Few-shot Classification}

\author{
\IEEEauthorblockN{Jinting Zhu\IEEEauthorrefmark{1},
Julian Jang-Jaccard\IEEEauthorrefmark{1},
Ian Welch\IEEEauthorrefmark{2},
Harith AI-Sahaf\IEEEauthorrefmark{2},
Seyit Camtepe\IEEEauthorrefmark{3} and
Aeryn Dunmore\IEEEauthorrefmark{1}}
\IEEEauthorblockA{\IEEEauthorrefmark{1}Cybersecurity Lab,Massey University, New Zealand\\
\IEEEauthorrefmark{1}Email: jzhu3@massey.ac.nz}
\IEEEauthorblockA{\IEEEauthorrefmark{2}School of Engineering and Computer Science, Victoria University of Wellington, Wellington, New Zealand}
\IEEEauthorblockA{\IEEEauthorrefmark{3}CSIRO Data61, Australia}}


%Telephone: (800) 555--1212, Fax: (888) 555--1212}


\maketitle

\begin{abstract}

When malware employs an unseen zero-day exploit, traditional security measures such as vulnerability scanners and antivirus software can fail to detect them. This is because these tools rely on known patches and signatures, which do not exist for new zero-day attacks. Furthermore, existing machine learning methods, which are trained on specific and occasionally outdated malware samples, may struggle to adapt to features in new malware. To address this issue, there is a need for a more robust machine learning model that can identify relationships between malware samples without being trained on a particular malware feature set. This is particularly crucial in the field of cybersecurity, where the number of malware samples is limited and obfuscation techniques are widely used. Current approaches using stacked autoencoders aim to remove the noise introduced by obfuscation techniques through reconstruction of the input. However, this approach ignores the semantic relationships between features across different malware samples. To overcome this limitation, we propose a novel Siamese Neural Network (SNN) that uses relation-aware embeddings to calculate more accurate similarity probabilities based on semantic details of different malware samples. In addition, by using entropy images as inputs, our model can extract better structural information and subtle differences in malware signatures, even in the presence of obfuscation techniques. Evaluations on two large malware sample sets using the N-shot and N-way methods show that our proposed model is highly effective in predicting previously unseen malware, even in the presence of obfuscation techniques.
\end{abstract}


% \fi
\IEEEpeerreviewmaketitle

\section{Introduction}
The security of nations and the privacy of individuals are highly dependent on the safety and dependability of electronic devices. The biggest risks in this sector often stem from unknown threats, where malware is a major contributor. To address these threats, cyber security professionals must have the ability to accurately and quickly detect potential hazards and identify ongoing attacks. This need has led to the development of advanced deep learning-based detection methods, which are essential for defence in the constantly evolving malware landscape. 

\textcolor{blue}{Malware detection refers to identifying whether a given piece of software or a file is malicious or benign. In the context of zero-day attacks, malware detection faces significant challenges because these attacks use unknown vulnerabilities and novel techniques that are not recognized by traditional detection systems. Detection methods often rely on known signatures, patterns, or behaviours that have been previously identified and documented. However, in the case of zero-day attacks, which exploit the unknown or unaddressed vulnerability is referred to as a zero-day vulnerability or zero-day threat, as these signatures or patterns are not available, making detection difficult.}

\textcolor{blue}{Given the popularity of Artificial Intelligence (AI) techniques, their application in detecting zero-day or unknown malware has become increasingly sophisticated, particularly with the integration of few-shot learning mechanisms. Few-shot learning, a subset of machine learning, enables AI systems to recognize new patterns or anomalies with minimal data examples. This is especially crucial for identifying zero-day threats, as these types of malware are often not well-represented in large datasets. By using few-shot learning, AI algorithms can quickly adapt to and recognize new, previously unseen forms of malware with very few examples, significantly enhancing the speed and efficiency of threat detection. This approach is a game-changer in cybersecurity, offering robust defenses against emerging and rapidly evolving digital threats}

Among them, many deep learning methods based on static and dynamic features have been proposed. Motivated by promising results in the use of AI, various feature-based detection models have been proposed, including using malware grayscale images \cite{kumar2022dtmic} and entropy graphs \cite{zhu2022few,chai2022data}. Although these static features have improved AI models in the detection of many known malware classes, they are vulnerable to new malware which is without compiled signatures \cite{9863062}. In addition, these existing methods are also vulnerable to slight changes in malware images, which also results in a decrease in the detection accuracy, most likely due to applying obfuscation techniques. To mitigate the influences of the obfuscation, \cite{salman2019denoising} and \cite{kumar2021ae} proposed enhancing the feature representation methods through an autoencoder. But they were still reported to be vulnerable to many types of unseen samples \cite{zahoora2022zero,mahdavifar2022effective}.

%Though various dynamic feature analysis has been proposed, the high cost associated with the behavioural analysis has been a barrier in most of the existing cybersecurity assessment tools on real systems \cite{sharma2022survey, furfaro2018cloud}. %For example, most software qualities are satisfied by adding some properties at a development stage, but this is going to have a costly impact, potentially leading to failure in the virtual environment \cite{furfaro2017using}. 

Due to the cost of dynamic feature analysis, more practitioners have defaulted to static feature analysis as it provides higher levels of efficiency\cite{saxe2015deep}. However, one of the disadvantages of the static feature analysis is that it is susceptible to inaccuracies due to polymorphic and metamorphic obfuscation techniques. Even with the utilization of AI tools in recent years, static feature analysis is still vulnerable to weak representations of embedded spatial features and thus seems to fail to capture core malware signatures \cite{jusoh2021malware}. One of the weaknesses of static feature analysis is that a single feature is not always enough to represent the complex interrelationships among malware. The limited abilities of single representations, such as the feature learned from the spatial hierarchies through a back-propagation algorithm, can not completely express the complex relationships across malware data because there is a non-linear relationship across the malicious code \cite{sharma2018efficacy}. Therefore, a model that can capture the relation of malware through feature embedding is needed. The artificial white noise can also influence the accuracy of a given detection model in a probability distribution of zero value in the grey image, such as no operation (NOP) obfuscation; \textcolor{blue}{ Entropy is employed to identify anomalies, unusual patterns, or irregular behaviours in data, often indicative of malicious activity. Malware and cyber-attacks typically manifest through such atypical data patterns, which are effectively discernible via entropy-based analysis. Additionally, this method demonstrates resilience against noise introduced by malware developers, enhancing its efficacy in cybersecurity applications}

Malware developers use obfuscation techniques to reduce the likelihood of detection \cite{zhu2015adaptive}. To help mitigate this potential threat, we posit that a robust feature relation can greatly help the task of malware detection \textcolor{blue}{, as malware functions that exist in a non-relation-aware context operate independently of the relationships or interactions among system components.}  We set out to examine relation-aware few-shot learning with robust features and to capture the underlying image regularity in malware samples in which a specific pattern links to the corresponding malicious function. By using feature relations \cite{hu2018relation,torabi2021strings}, we can uncover groups of correlated malware with common features \cite{xu2016hadm, kim2022obfuscated, zhou2020siamese}. Sharma et al. \cite{sharma2018efficacy} found that the underlying physical processes behind converting malware binaries to images are highly non-linear \cite{sharma2018efficacy}, and this correlation exists between malware instances. Other works tend to focus on either the correlation between the malware samples \cite{chai2022dynamic} or only on an individual robust feature for the malware samples.
In this study, we provide a novel insight into how detection mechanisms are constructed using a relation-aware Siamese embedding in the context of few-shot learning to classify unseen, obfuscated malware samples. %We aim to explore the framework best suited to the real scenario of malware detection. 
The major contributions of this paper are:

\begin{itemize}

\item Our proposed solution involves utilizing entropy-based features to train our model, rather than relying solely on traditional malware image features. The use of entropy-based features allows for a more comprehensive capture of the distinctiveness and structural information of the malware. This leads to improved accuracy when identifying different malware signatures and discovering similarities in obfuscated malware. This innovative approach significantly contributes to the effective differentiation of malware classes.

   % \item We propose to use a novel Siamese Neural Network (SNN) with denoising autoencoders. The use of SNN allows our learning model to accurately classify malware classes even if only a limited number of malware samples are available. In addition, the use of a denoising autoencoder embedded with a relation-aware module at each branch of SNN captures the semantic differences of the features from a pairwise malware that may have complex nonlinear relationships.

\item We propose the implementation of a cutting-edge Siamese Neural Network (SNN) combined with denoising autoencoders. The utilization of SNN enables our machine learning model to efficiently classify malware classes, even when limited samples are available. Moreover, the integration of denoising autoencoders with a relation-aware module in each branch of the SNN enables us to effectively capture the semantic differences and the complex nonlinear relationships between features, in a pairwise malware comparison.

   % \item  Our proposed model replaces the distance scores as a measure to capture the semantic difference in different malware classes but uses relation-aware embeddings which are better suited to output more accurate similarity probabilities across different malware samples.

   \item Instead of relying on traditional distance scores, our model incorporates relation-aware embeddings to output more precise similarity probabilities between various malware samples. This innovative approach enhances the accuracy of our learning model and enables it to distinguish between different malware signatures more effectively.

  %  \item Extensive evaluation and analysis based on two sets of relatively large malware samples on the N-shot and N-way methods show that our proposed model is highly effective in predicting unseen malware samples (i.e., zero-day malware attack) even though they are disguised with obfuscation techniques.

  \item Our proposed model has undergone thorough evaluation and analysis using two substantial sets of malware samples and the N-shot and N-way methods. The results demonstrate that our model is highly efficient in identifying zero-day malware attacks, even those modified by obfuscation techniques.

\end{itemize}

The next section begins with related work in Section \uppercase\expandafter{\romannumeral2}.  Section \uppercase\expandafter{\romannumeral3} introduces the preliminary materials in the obfuscation technique. Section \uppercase\expandafter{\romannumeral4} provides the method for constructing our model and the details of our network architecture. In Section \uppercase\expandafter{\romannumeral5}, we provide an ablation experiment to clarify our model's ability and provide multiple analyses for hyperparameters. Finally, this paper concludes in Section \uppercase\expandafter{\romannumeral6}, which provides a summary, conclusion, and potential directions for future work.

\section{Related Work}
\subsubsection{Relational exploring for malware detection} 
 RevealDroid \cite{tang2022android} simultaneously adopted multiple types of features to train their detection model for Android malware. This method declares that it supports the resistance of four obfuscation techniques: API reflection obfuscation, class name obfuscation, array encryption, and string encryption. However, with the limited capability of the model as trained and tested on the seen classes, it is difficult to apply in a real scenario where unknown attacks are common. Xu et al. \cite{xu2021malicious} explored the structural and semantic relations in Android applications through the entity feature combined with matrices and meta-paths. Moreover, the imbalanced property of malicious behavior exists universally in this field.  Zheng et al. \cite{zheng2020learning} applied a meta-learning approach to a classification task of encrypted traffic and to classify unseen categories based on a few labeled samples. Han et al. \cite{han2019maldae} analyze the underlying correlation given by the explainable framework between the dynamic and static API call sequences of malware in order to construct the hybrid feature vector space. Nikolopoulos et al. \cite{nikolopoulos2022behavior} constructed the System-call Dependency Graphs obtained through the dynamic taint analysis over the execution of a program that exploited the valuable structural characteristics of the augmented graphs.  Mpanti et al. \cite{mpanti2018graph} generated the graph given by degrees and the vertex-weights by utilizing the functionality of system calls to extend the representation of malicious behaviors. Above all, generalization on unseen malicious attacks without depending overly on data size has proven challenging.
\begin{figure*}
\centering
  \includegraphics[scale=0.35]{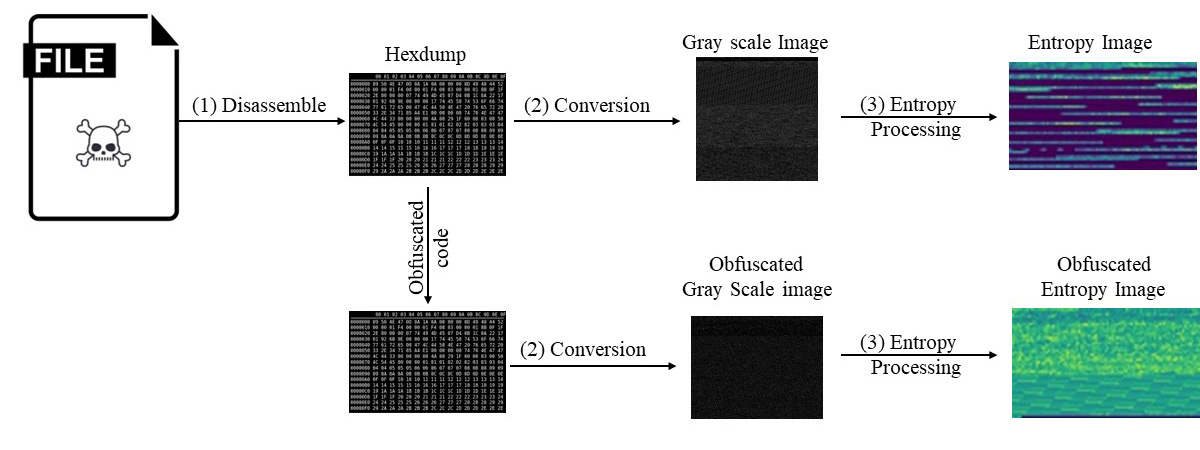}
  \caption{Feature Processing for Entropy Image}
  \label{fig:disassembler}
\end{figure*}

\subsubsection{Malware classification with the feature extraction}

%Although dynamic features consisting of API call sequences and opcode sequences \cite{dinh2019behaviour} are logic and reasonable in the term of coding function than the static feature for a person engaged in cybersecurity area, it is time consuming and have potential threaten for the operation system. Moreover, the operation on the virtual environment could result in the ..it is also possible not be triggered by the virtual environment. While the bytecodes are easily to be understood by the algorithms,
The cybersecurity community has explored automated malware behavior analysis in the last decade. Many detection mechanisms have been proposed to prevent attacks on individual and national data. The cognitive mechanism to understand the malware characteristics in the semantic information involves the dynamic and static features \cite{han2019malinsight, rabbani2020hybrid, noor2018countering, morato2018ransomware, wang2019mobile}, though the hybrid features \cite{tong2017hybrid, damodaran2017comparison, yoo2021ai} are also considered in some cases. 

N-grams have been used as features in several works and are one of the most common feature types for static analysis. Byte n-grams are particularly attractive since they require no knowledge of the file format and do not require any dynamic analysis. In this manner, one could potentially learn information from both the headers and the binary code sections of an executable \cite{raff2018investigation, johnson2022ensemble}. This approach is similar to the research proposed by Kang et al. \cite{kang2013android}, which captured similar information through the bytecode frequency. Additionally to the use of n-gram features for the static analysis, Nataraj et al. \cite{nataraj2011malware} first proposed a technique to extract pixel features from the grayscale images translated from the raw bytecode PE files of malware. Motivated by these grayscale features, Bakour et al. \cite{bakour2021visdroid} proposed a hybridized ensemble approach using both local and global features as a voter to make a decision in an ensemble voting classifier. Although they take the overfitting problems related to the imbalanced dataset into account, in addition to the small number of malware samples, they ignore the effect on performance caused by the polymorphic obfuscation. In such a situation, the texture of the malware images could be intentionally changed. Without this consideration, performance in existing malware detection research can achieve high accuracy on some datasets, but it cannot effectively detect obfuscated variants of malware, and as such the effectiveness of these methods drops dramatically. 

Jeon et al. \cite{jeon2021hybrid} also proposed a hybrid scheme for malware detection that extracts the static features of the opcode sequence in the static feature classification step. To overcome the shortages caused by static in the obfuscated malware, they also take the  dynamic features into account in the next step and execute the files in a nested virtual environment. Although the improved approach enables classifying the obfuscated malware based on hybrid features, most of the existing cybersecurity assessment tools act on real systems, incurring  high costs and risk \cite{sharma2022survey, furfaro2018cloud} and potentially leading to failure \cite{furfaro2017using} in the virtual environment. 

\subsubsection{Denoising autoencoder for malware detection}
One method of unsupervised learning within the research consists of an autoencoder architecture \cite{de2018malware, wang2019effective, kim2018zero}, which can learn the reconstructed features from noised samples. Alahmadi et al. \cite{alahmadi2022mpsautodetect} aim to extract meaningful features with the use of stacked convolutional denoising autoencoders (CAE) because the obfuscation and the complex nature of these malicious scripts causes bias in feature selection. Sandra et al. \cite{sandra2021bm3d} regarded the adversarial examples as noise that assists the malware samples in evading detection based on a deep learning model. To improve malware defenses, they aim to eliminate most noise in malware images through the denoising autoencoder. In addition, Salman et al. \cite{salman2019denoising} proposed an unsupervised deep learning-based model based on denoising autoencoders that de-anonymizes the mutated traffic to detect a malicious attack using obfuscation techniques which, when applied to the traffic's statistical characteristics, cause a misclassification.

\begin{figure}
\centering
  \includegraphics[scale=0.20]{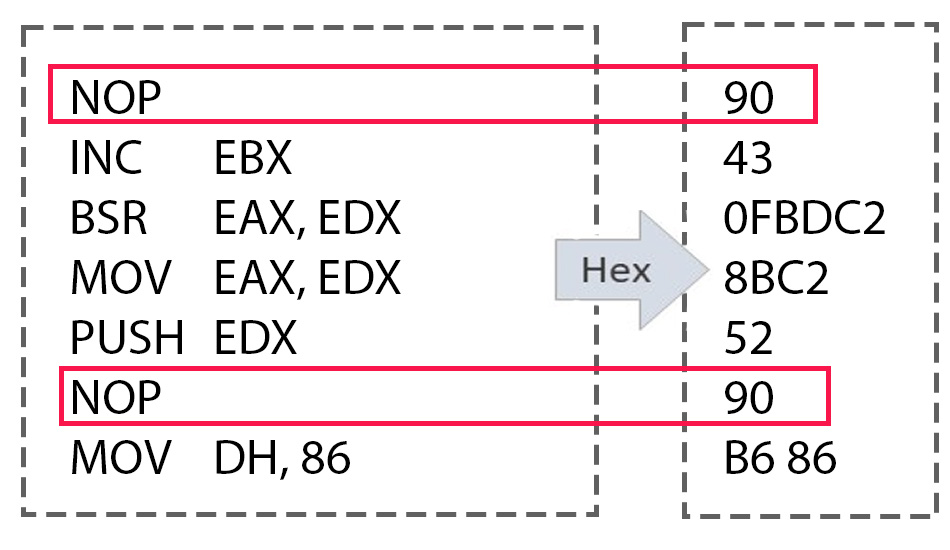}
  \caption{Obfuscated Code of No Operation (NOP) Insertion}
  \label{fig:nop}
\end{figure}
\section{Obfuscation Materials}

\subsection{Bytecode-based Junk code for Obfuscation}
Junk code obfuscation is a technique which makes executable files impossible to decompile into source code, or impedes the ability to understand a decompiled program, even while the original semantic function is preserved. Jien-Tsai et al. \cite{chan2004advanced} proposed the technique to intentionally introduce syntactic and semantic errors, called junk code, into a decompiled program such that the  program would have to be debugged manually, resulting in a bytecode, signature-based anti-virus generating an error or failing in the task of malware detection. 

Obfuscation techniques can be applied to benign software to prevent the benign application from being implanted with malicious functions or intentionally modified for other interests. Alternatively, the obfuscation technique can also be exploited into malware to defend against static analysis methods and top-rated anti-malware products. Malware authors employ polymorphic and metamorphic obfuscation
techniques, such as No Operation (NOP) insertion, altering the control flow into the source function to evade anti-malware solutions which use the opcode sequence as the malware signature \cite{faruki2014evaluation,chan2004advanced}. Under this situation, a variant may be included with an arbitrary number of additions, modifications, or deletions to the code by inserting malicious functions, so as to maintain the original semantic information. Specifically, a malware variant is represented as: $ M_{app} = m_1,m_{mod_2}, m_3,m_{add_1},m_{add_2},...,m_n$, where $m_{add_1}$, $m_{add_2}$  are inserted codes and $m_{mod_2}$ is the corresponding code modification of $m_2$ while still maintaining the intended functionality. Evolutionary possibilities of malware variants are more pronounced in opcode or hexdump \cite{stibor2010study}. To visualize this procedure, Fig. \ref{fig:disassembler} illustrates the connection between machine instructions, opcodes, and the hexdump. The hexdump is usually a common representation for a disassembled analysis, also known as the static representation of an executable file (or data in general), related to machine instructions or termed opcodes. It is a technique that aims to analyze several types of files, including execution files, shared libraries, object files, etc. The easiest way to obfuscate malware samples is via the bytecode, which is considered as an interpreter executes. It can also be compiled into machine code (opcode) for the target platform. We implemented the operation on the bytecode with the hexdump of malware samples in the datasets to create our synthetic dataset used in this paper.

%\begin{table}[]
    %\centering
       %\caption{Relation between machine instructions, opcodes and the consequence %of  hexdump}
%\begin{tabular}{ |p{2.0cm}|p{2.0cm}|p{3cm}|}
%\hline
%\multicolumn{3}{|c|}{Dalvik  bytecode Examples} \\
%\hline
%Opcode (hex) & Opcode name &  Description  \\
%\hline
%00 & nop & Perform no operation \\
%01 & move vx,vy  & Moves the content of vy into vx. Both registers must be in the %first 256 register range.\\
%07 &move-object vx,vy & Moves the object reference from vy to vx.\\
%0A &move-result vx & Move the result value of the previous method invocation into %vx.\\
%...   & ... & ...\\
%0F & return vx & Return with vx return value \\
%\hline
%\end{tabular}
   % \label{tab:my_label}
%\end{table}

\subsection{Obfuscation of No Operation (NOP) Insertion}
No-op instruction (NOP) obfuscation, as shown in Fig \ref{fig:nop}, is used to waste the CPU execution cycle, where NOP is an assembly language instruction that does nothing. An obfuscated malware sample synthesizes the original Android samples with NOP instruction, which is randomly added into the disassembled methods while preserving the semantic information \cite{faruki2014evaluation, you2021comparative}. It is possible to evade malware detection solutions employing opcode sequences as malware signatures, through repeats of randomly inserting the bytecode of NOP multiple times at ambiguous positions in the hexdump form of a malware sample. 

%\begin{figure}
%\centering
 % \includegraphics[scale=0.7]{junk-code.png}
  %\caption{Obfuscation of Registers Allocation}
  %\label{fig:registers}
%\end{figure}

\section{Methodology }

Our approach involves a few-shot classification task using a Siamese denoising autoencoder against junk code obfuscation. The details of the proposed scheme are presented in this section. The evolution of malware samples often has a certain inheritance correlation with the key functions of previous versions, or the key functions are packaged with multiple obfuscation techniques to evade detection. The core of its detection is to mine such a property that it will be able to detect new versions that exist or evolve in the future. A generic few-shot learning method establishes a metric embedding trained through the support and query samples. In this work, the relationships are established by a Siamese relation-aware feature embedding, as shown in Fig \ref{fig:few-shot}. With the embedding learned through the fully connected layer (FC) or the convolutional layer (CNN), an input feature can be projected onto the embedding to calculate the relation score between the support sample and the query sample. A relation-aware embedding trained in this manner can predict samples in the untrained class through the query samples. 

\subsection{The Entropy Feature Conversion}
\begin{figure}[th!]
    	\centering
    	\subfloat[Bitman Entropy Pattern] {\includegraphics[width=0.17\textwidth]{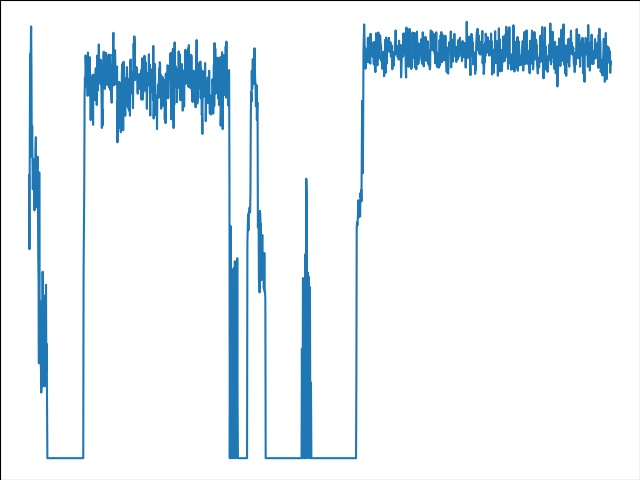}} \hfil
    	\subfloat[Bitman Entropy Image] {\includegraphics[width=0.19\textwidth]{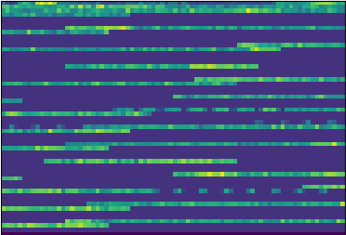}} \hfil
    	\subfloat[Upatre Entropy Pattern] {\includegraphics[width=0.17\textwidth]{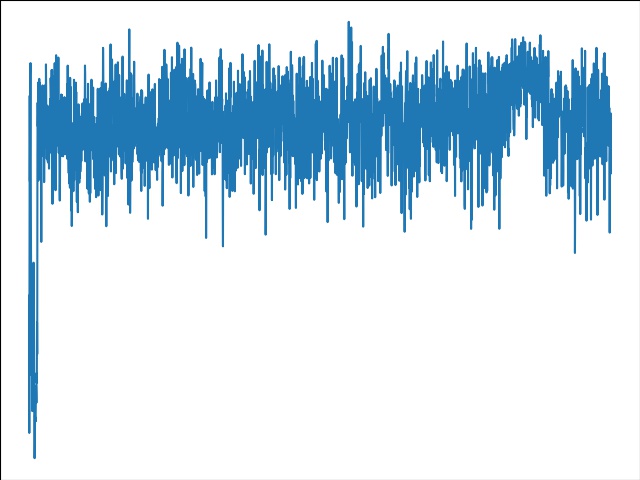}} \hfil
    	\subfloat[Upatre Entropy Image] {\includegraphics[width=0.19\textwidth]{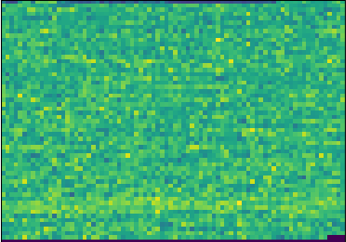}}
    	\caption{Entropy pattern  corresponding to entropy image}
    	\label{fig:entropy_image}
\end{figure}

\begin{figure*}
\centering
  \includegraphics[scale=0.52]{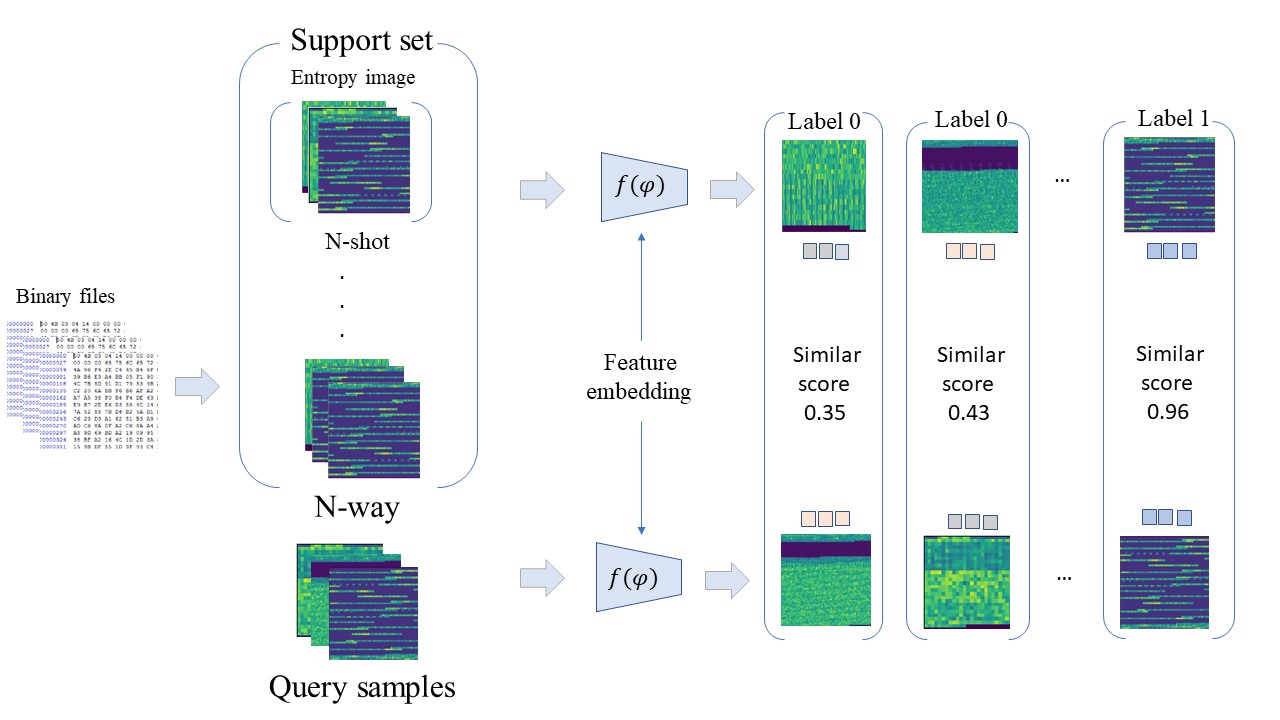}
  \caption{The diagram of Few-shot learning}
  \label{fig:few-shot}
\end{figure*}

%\begin{figure}[th!]
    %	\centering
    %	\subfloat[Original Bitman.a1] {\includegraphics[width=0.18\textwidth]{0c7d0ad0454b780997f5d6856cf87920e9fe667634bd8ccab53177d5f32de5a2.jpg}} \hfil
    %	\subfloat[Obfuscated Bitman.a1]
    %	{\includegraphics[width=0.18\textwidth]{ob_0c7d0ad0454b780997f5d6856cf87920e9fe667634bd8ccab53177d5f32de5a2.jpg}} \hfil
    %	\subfloat[Original Bitman.a1]
    %	{\includegraphics[width=0.18\textwidth]{EIMAGE_01.png}} \hfil
    %	\subfloat[Obfuscated Bitman.a1]
    %	{\includegraphics[width=0.18\textwidth]{Ob_EIMAMGE_0.png}}
    %	\caption{The appearance comparison between original  and obfuscated in gray scale image and entropy Image}
    %	\label{fig:entropy}
%\end{figure}
Entropy feature conversion aims to measure the uncertainty distribution of the information. Moreover, Vidya et al. \cite{vidya2019entropy} argue that the entropy feature has better feature representation since it exhibits higher uniqueness and entropy values are incorporated from local regions to add extra information content to these images. In the context of the malicious code, we could build the malicious pattern with characteristics of entropy information calculated from the binary pattern. In \cite{gibert2018classification,canfora2016hmm}, the authors divided bytecodes into blocks of fixed size, and then computed the Shannon entropy for each block. We attempt to further their achievements in image recognition using the entropy value converted from a binary pattern \cite{wang2019mobile}. We assume that the entropy information can be converted to an entropy image. This approach preserves the connected information of each entropy sequence in the context of the raw bytecode data, as Fig. \ref{fig:entropy_image} shows. The entropy images are comprised of global and local entropy information that can be calculated with the Shannon entropy equation below:
\begin{equation}
    Ent = -\sum_{i}\sum_{j}M(i,j)logM(i,j)
\end{equation}
where $M$ is the probability of an occurrence of a byte value. The Shannon entropy equation obtains the minimum value of 0 when all bytecode of the malware have no changes. Alternatively, the maximum entropy value of byte value 8 is obtained when all the bytecode is different. If specific information of bytecode occurs with high probabilities, the entropy value will be smaller. 
To visualize the entropy pattern that reflects the pattern of entropy information and the intrinsic connection of each sequence of raw bytecode, we built a grey-scale matrix through Equation \ref{grayscalematrixeq}, and the entropy values are then concatenated into a stream of values that can form an entropy sequence.

\begin{equation}\label{grayscalematrixeq}
    P_{(i,j)} = 2^{ent}-1
\end{equation}
where $ent$ denotes the entropy value of the bytecode block $(i,j)$ and $P_{(i,j)}$ means that the gray pixel value is in the range of [0, 255]. The entropy images are concatenated with the full entropy sequences sequentially to generate an $105 \times 105$ entropy image. However, the sizes of bytecodes are different, and we fix the width value to resize the image size to $105 \times 105$.

\subsection{Denoising AutoEncoder}
%To explore the denoising autoencoder against junk code obfuscation, we investigated this network which varying field has been involved with. Its applications have been widely applied in various fields using denoising autoencoder, such as image inpainting \cite{xie2012image}, signal denoising\cite{zhu2019seismic} and music removing \cite{zhao2015music}, in an unsupervised manner. %
In cybersecurity research, denoising encoders have been widely used in recent years \cite{de2018malware,shamsuddin2020transforming,d2020malware}.This can be attributed to the denoising autoencoder (DAE) network's aim to reconstruct the data to its original characteristics or its uncorrupted version, without the noise. This noise can be considered to be Gaussian noise, music \cite{zhao2015music}, occluded faces \cite{gorgel2019face}, or even junk code.

A typical denoising autoencoder can be stacked using multiple convolutional layers consisting of encoders ($E^{\frac{M}{2}}$) and decoders ($D^{\frac{M}{2}}$) in varied depths. Specifically, the first $M/2$ hidden layers encode the input as a new representation, and the last $M/2$ layer decodes the representation in the latent embedding to reconstruct the input. We can define this equation as:

\begin{equation}
z_i^{(m,v)} = a(W_{ae}^{m,v}z_i^{m-1,v}+b_{ae}^{(m,v)})
\end{equation}
where $z_i^{(m,v)} \in  \mathbb{R}$ and $d_{m,v}$ is the  number of nodes, $\{W_{ae}^{m,v},b_{ae}^{(m,v)}\}$ is the parameter set for all layers with 
$M+1$ being the number of layers of our network while $a(\cdot)$ is a nonlinear activation function. The feature matrix is $X^{(p)} = [x_1^{p},x_2^{p}, ,...,x_3^{p}],  \in \mathbb{R}^{d_{p \times n}}$ for one of the sub-networks. Meanwhile, the corresponding reconstructed representation is denoted as:
\begin{equation}
Z^{(M,v)} = [z_1^{(M,v)}, z_2^{(M,v)}, ... , z_3^{(M,v)}]
\end{equation}
where $z_i^{(M,v)}$ is the reconstructed representation for the $i$th sample in one sub-network. Thus, the low-dimension representation $Z^{(\frac{M}{2},v)}$ is obtained by the following reconstruction loss with mean square error (MSE):

\begin{equation}
     \mathcal{L}_{mse} = \frac{1}{2n}\sum^n_{i=1}\|X^{(p)}-Z^{(M,v))}\|^2_F
\end{equation}
where $X^{(p)}$ and $Z^{(M,v))}$ is the noise sample inputted and the original samples, respectively.

\subsection{Siamese-based Denoising AutoEncoder}

Our model contains a Siamese neural network (SNN) that is commonly used in contrastive learning. The SNN model is useful on pairwise identity verification, and it takes as input two images in order to identify a meaningful distance between the representations of those two images. Each of the sub-network is parameterized by the shared weights and bias $\{W_{ae}^{m,v},b_{ae}^{(m,v)}\}$, performed on both input images whether or not they are the same.

With the connection part of encoders ($E^{\frac{M}{2}}$) and decoders ($D^{\frac{M}{2}}$), we extracted the features out of the latent embedding optimized by the siamese denoising autoencoder. A relation-aware function then calculates the correlation between the two input images. This correlation of the pair of images for the latent features in the last fully connected layers within the encoder ($E^{\frac{M}{2}}$) can be denoted as:
\begin{equation}
    d_\varphi(z_i,z_j) = \|g_\phi(z_i^{(\frac{M}{2},v)})-g_\phi(z_j^{(\frac{M}{2},v)}) \|_F^2
\end{equation}
where notation $d_\varphi(z_i,z_j)$ is represented as $d_\varphi$ and $d_\varphi$ denotes whether the $z_i$ and $z_j$ are similar. When this is similar the $d_\varphi$ is close to one, or zero otherwise. The contrastive loss to calculate their relationship is computed by:

\begin{equation}
    \mathcal{L}\sum^p_{i=1}L(\varphi,(z_i^{(\frac{M}{2},v)},z_j^{(\frac{M}{2},v)})^i)
\end{equation}

\begin{equation}
    L(\varphi)=(1-y_i )L_s (d_\varphi )+y_iL_d (d_\varphi)
    \label{eq:logsitic}
\end{equation}
The classification ability depends on the performance of  $d_\varphi$ calculated by Eq. \ref{eq:logsitic}, which linearly represents the Euclidean distance. 

\subsection{Siamese in Relation-Aware Embedding for  Malware Classification}
Strategies such as \cite{vasan2020imcfn, kumar2021mcft} only take the independent malware signature into account, or rarely consider the non-linear relationship \cite{hsiao2019malware, sison2021calculating} in the contextual difference between all malware pairs. Moreover, malware classification often struggles in data-poor problems where the underlying structure is characterized by organized but complex relations. This is especially true for the interaction of functions of coding between malware samples.

We propose a relation-aware Siamese denoising autoencoder that enhances the conventional SNN with relational semantic information. We jointly learn new representations in the non-linear relationships for $\mathcal{C}$ that are assumed to represent the concatenation of feature maps in depth that can explore a learnable rather than fixed metric, or non-linear rather than linear classifier \cite{sung2018learning,zhang2020reinforced,cheng2021learning}.
\begin{figure*}
\centering
  \includegraphics[scale=0.45]{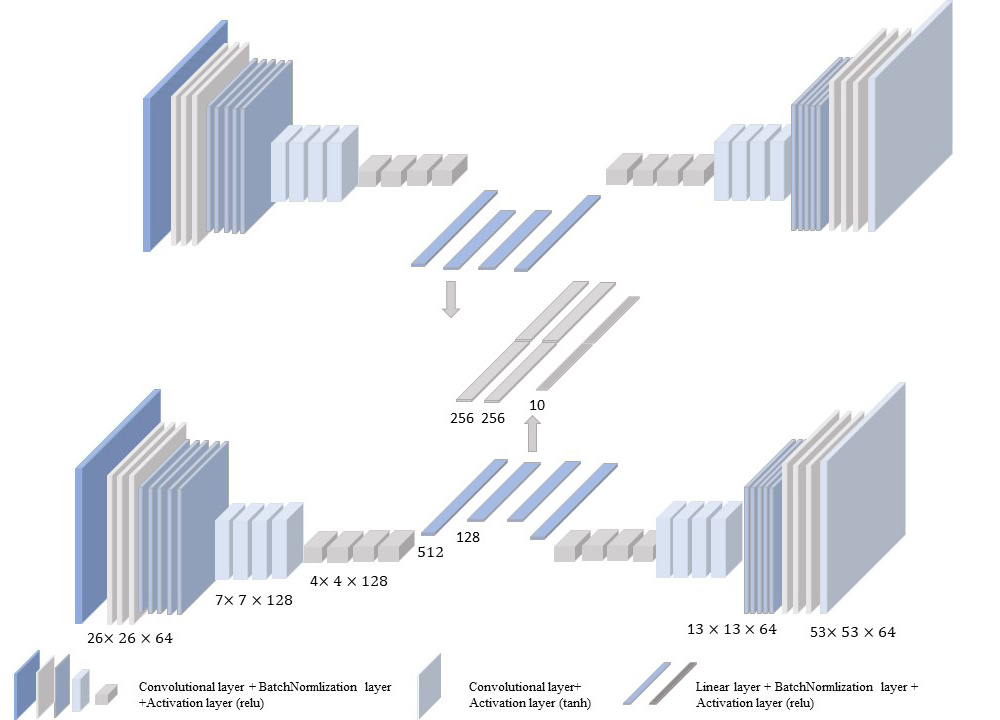}
  \caption{The diagram of our model architecture}
  \label{fig:boat1}
\end{figure*}

Modeling the relations between the instance pairs has been shown to improve the results of classification, and these latent features can be explored by this relation module to determine if they are similar using the relation classifier.

\begin{equation}
   r_{i,j} =  \mathcal{C}(f_\varphi(z_i^{(\frac{M}{2},v))}, f_\varphi(z_j^{(\frac{M}{2},v))}), \ \ \ \  i = 1,2,...,C
\end{equation}
where the projected features, $z_i^{(\frac{M}{2},v)}$ , $z_j^{(\frac{M}{2},v))}$ of the support sample and query sample, are aimed to be combined with each other. These are 
then fed through the sub branch of the relation-aware siamese neural network $f_\varphi$. 

\begin{equation}
   \mathcal{L}_{r}=\sum^i_{i=1}\sum^j_{j=1}  (r_{i,j} - y_{i,j})^2
\end{equation}
Given the set of latent features $\mathbf{z}$ = $\{z_1,z_2,...,z_n\}$ from the latent embedding for a malware sample and $y_{i,j}$ denotes the one-hot encoding for the ground-truth label of malware family. Finally, the objective function for few-shot learning is: 
\begin{equation}
\mathcal{L} =\mathcal{L}_{r}+\lambda \mathcal{L}_{mse}
\end{equation}
where the value of hyperparameter $\lambda$ settled as 0.7.

\subsection{Network Architecture}
In this section, we describe the specific parameters in the architecture of our model. First, we apply convolutional layers at the front of the CAE. One advantage of adding a convolutional network is that it is highly invariant to translation, scaling, tilting or other forms of deformation. Furthermore, the fully-connected layers at the front part are one of the major causes of for increasing the number of parameters. We put convolutional layers first and stack the fully-connected layers on top of convolutional layers for ease of the reduction to the target dimension $\mathbb{R}^{d}$.

The block diagram of the proposed image compression based in Fig. 5 shows the architecture of the layers, stacking up to 4 convolutional and batch-normalization layers. The only preprocessing step before the CAE network consists of normalizing the entropy pattern to [−1, 1] by calculating the value of mean and standard deviation of 0.52206 and 0.08426 respectively in the VUW dataset. The size of the input is denoted as $H \times W \times C$, where $C$ represents the number of colour components. We considered $C = 1$ for the entropy images due to the conversion to grey scale images.

%The analysis transform and synthesis transform have symmetric networks, apart from using convolutional and deconvolutional filters, respectively. 

%\textcolor{blue}{RC,RL...- It is difficult to interpret the Figure 5. Illustrate the main architecture to explicitly reveal the main method such as relation-aware embedding.
%- The notation in algorithm 1 should be matched in the main text (e.g., RC, RL).
%}

%\textcolor{blue}{Algorithm 1}

In term of network design, a siamese neural network is more robust to class imbalance \cite{bedi2020siam} as it focuses on learning embeddings (in the deeper layer) that place the same classes/concepts close together in order to learn semantic similarity. We have also considered the term of the receptive field of the topmost feature layer in each sub-branch, because if the receptive field is too large it means that there are too many layers, and the risk of network overfitting will increase. Meanwhile, the entire network is difficult to converge.
%The whole model  architecture is included in the encoder part, decoder part, and the relation part that are calculated for the classification task, as shown in Figure \ref{fig:few-shot}.%

\textcolor{blue}{Algorithm1}

By considering the Siamese theory of conception, the encoder part of our model is designed with 4 convolution layers in which each layer is followed by a batch normalization layer and relu activation layer. All convolutional filters have size of 3 $\times$ 3 and use a stride of 2 for the encoder part. The order of the CNN layers in the decoder part is opposite to the encoder part. A max pool layer is added after the first convolution layer. Two linear layers are used to improve the classification performance with the extracted features through the encoder part, and it is the key module for relational embedding. The extracted features have are robust against the NOP obfuscation as they have been training with autoencoder module. We further compare the performance with the different dimensions of the linear layer and find that the output dimension of 256 for the first linear layer has better performance than that with a dimension of 128. Finally, the sigmoid activation function is taken as the outcome function for relation label 1 or 0.

\begin{algorithm}[hbt!]
		\SetKwInOut{Input}{Input}
		\SetKwInOut{Output}{Output}
		
		\Input{\textit{Folder\_root : f, Class\_num : cc, Num\_per\_class : np, Batch\_num : bn, obfuscated samples $x_b^1,x_b^2$ and original samples $x_o^1,x_o^2$, Reconstruction loss : RC,  Relaton loss : RL   } }
		\Output{Relation score for each pair}
		
Dataloader = FewShotTask(f, cc, np, bn)\\
		
 for  episoded in range(EPISODE) \\
   \ \ \ \ \   $(x_b^1,x_b^2)$  =  Dataloader.iter() \\
	   \ \ \ \ \     $en_1,en_2$ = Siamese\_EncoderNetwork$(x_b^1,x_b^2)$ \\
		      \ \ \ \ \                 $dn_1, dn_2$ = Siamese\_DecoderNetwork($en_1,en_2$) \\\
        \ \ \ \ \ Pairs $(r_n)$ = Feature\_concatenation(($en_1$,$en_2$)) \\
		    \ \ \ \ \ Score ($s_z$) = Siamese\_RelationNetwork($(r_n)$)     \\  
		     \ \ \ \ \ $Loss_1$ =  RC ([$dn_1,en_1$],$[dn_2,en_2$])  \\
        \ \ \ \ \ $Loss_2$ = RL ($s_z$)  \\
        \ \ \ \ \ $Loss$ = 0.7*$Loss_1$ + $Loss_2$  \\

		     \ \ \ \ \ $Loss$.backward()\\
		     \ \ \ \ \ Optimizer.step()
		
		\caption{Pseudocode for Our Proposed Algorithm}
		\label{alg:alg1}
\end{algorithm}

\section{Experiments}
We evaluated the performance of our proposed model on two malware datasets according to the few-shot learning mechanism. We also conducted the ablation experiments to analyze the effect of various parameters, such as pooling methods, layer dimension, and hyperparameters. The experiments were conducted on equipment consisting of 32GB RAM, Nvidia Geforece RTX 2070(8GB), and Intel i7-9700 CPU@ 3.00GHz. 

\begin{figure*}[th!]
    	\centering
    	\subfloat[Original Entropy Image] {\includegraphics[width=0.22\textwidth]{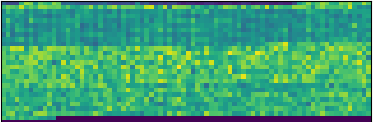}} \hfil
    	\subfloat[Frequency 200 with Nop] {\includegraphics[width=0.22\textwidth]{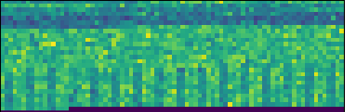}} \hfil
    	\subfloat[Frequency 400 with Nop]
    	{\includegraphics[width=0.22\textwidth]{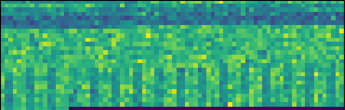}} \hfil
    	\subfloat[Frequency 600 with Nop]
    	{\includegraphics[width=0.22\textwidth]{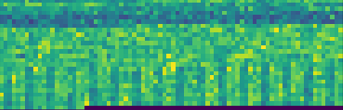}} 

    	\caption{The appearance comparison between original images and obfuscated images}
    	\label{fig:entropy}
\end{figure*}

\subsection{Description of dataset}
\paragraph{VUW dataset} This dataset utlized VirusShare\footnote{VirusShare. https://virusshare.com/} to collect malware samples. With the VUW dataset, we collected a total number of 1,048 samples from 11 families/classes of ransomware, each consisting of a varying number of examples, as listed in Table \ref{table:malware-family}. This dataset intuitively reflects the distribution of the data in real situations, as some classes, e.g., Petya and Dalexis, are largely outnumbered by the other classes, e.g., Zerber, as shown in the third column of Table \ref{table:malware-family}.

\paragraph{Malimage dataset}
The details of the second malware dataset, Malimage, published in \cite{nataraj2011malware} are listed in Table \ref{table:malimage}. There are 9,314 instances spread over 25 families in which the width and height of each malware image differ between families. Nataraj et al. \cite{nataraj2011malware} fixed the width to a certain length according to the file size and the bytecode sequence. Observed in the Table \ref{table:malimage}, the number of samples is greater than the VUW dataset. The malware samples have distinctive image textures across the different malware families. 

\begin{table}[h]
		\centering
  \caption{Details of the VUW ransomware dataset}
		{\footnotesize
			\begin{tabular*}{0.48\textwidth}{@{\extracolsep{\fill}}|lcc|}
				
				\hline
			Family name & Instances & Ratio (\%) %& Packed  (\%) 
				\\ \hline
				
				Bitman & 99 & 9.45 %& 19.2
				\\
				Cerber & 91 & 8.68 %& 30.8
				\\
				Dalexis & 9 & 0.86 %& 11.1
				\\
				Gandcrab & 100 & 9.54 %& 1.00 
				\\
				Locky & 96 & 9.16 %& 35.4
				\\
				Petya & 6 & 0.57 %& 0.00 
				\\
				Teslacrypt & 91 & 8.68 %& 28.6 
				\\
				Upatre & 18 & 1.72 %& 5.60
				\\
				Virlock & 162 & 15.46 %& 8.00 
				\\
				Wannacry & 178 & 16.98 %& 49.4 
				\\
			Zerber & 198 & 18.89 %& 10.1 
				\\
				\hline
			
			\end{tabular*}
 \label{table:malware-family}

		}
\end{table}

\begin{table}[h]

		\centering
  \caption{Details of the Malimage dataset}
		{\footnotesize
			\begin{tabular*}{0.48\textwidth}{@{\extracolsep{\fill}}|lcc|}
				
				\hline
			Class name & Family & Instances %& Packed  (\%) 
				\\ \hline
				
				Worm & Allaple.L & 1591 %& 19.2
				\\
				Worm & Allaple.A & 2949 %& 30.8
				\\
				Worm & Yuner.A & 800 %& 11.1
				\\
				PWS & Lolyda.AA 1 & 213 %& 1.00 
				\\
				PWS & Lolyda.AA 2 & 184 %& 35.4
				\\
				PWS &   Lolyda.AA 3 & 123 %& 0.00 
				\\
				Trojan & C2Lop.P & 146 %& 28.6 
				\\
				Trojan & C2Lop.gen!g  & 200 %& 5.60
				\\
				Dialer & Instantaccess & 431 %& 8.00 
				\\
				TDownloader & Swizzot.gen!I & 132 %& 49.4 
				\\
			TDownloader &  Swizzor.gen!E & 128 %& 10.1 
				\\
				Worm & VB.AT & 408 %& 10.1 
				\\
				Rogue & Fakerean & 381 %& 10.1 
				\\
				Trojan & Alueron.gen!J & 198 %& 10.1 
				\\
				Trojan & Malex.gen!J & 136 %& 10.1 
				\\
					PWS & Lolyda.AT & 159 %& 10.1 
				\\
				Dialer & Adialer.C & 125 %& 10.1 
				\\
					TDownlaoder & Wintrim.BX & 97 %& 10.1 
				\\
					Dilaer & Dialplatform.B & 177 %& 10.1 
				\\
					TDownlaoder &  Dontovo.A & 162 %& 10.1 
				\\	TDownlaoder &  Obfuscator.AD & 142 %& 10.1 
				\\	Backdoor & Agent.FYI & 116 %& 10.1 
				\\
					Worm:AutoIT & Autorun.K & 106 %& 10.1 
				\\
					Backdoor & Rbot!gen & 158 %& 10.1 
				\\
					Trojan & Skintrim.N & 80 %& 10.1 
				\\
							\hline
				
 \end{tabular*}
\label{table:malimage}
		}
\end{table}

\begin{table}[h]
\centering
\caption{Model learning parameters and their values}
\begin{tabular}{|l l | l |} 

 \hline
 Methods & Values & Description \\ [0.4ex] 
 \hline
 rescale & 1./255  &   Resizing an image by a given  scaling \\ 
 
 &   &  factor. \\ 
learning rate & 1e-02 & Epsilon for ZCA whitening. \\
batch size & 19/10-15 & 1-shot for 19 batch size; \\

  &   &  5-shot for 15 batch size   \\
rotation\_degree & 90/180/270 &  Setting degree of range for random   \\

& &  rotations.\\

mean & 0.52206 &  the average value in VUW 
\\
std& 0.08426 &  the standard deviations in VUW 
\\
$\lambda $& 0.7  &  the hyperparamter for the loss function

\\

 \hline
\end{tabular}

\label{table:parameters}
\end{table}

\begin{table*}[]
    \centering
 \caption{Accuracy scores (\%)  tested on two dataset}
 \setlength\doublerulesep{0.4pt}
 
\begin{tabular}{ p{3.5cm} p{2cm} p{2cm} p{2cm} p{2cm}}
\toprule[1pt]\midrule[0.3pt]
 %& 2-way 1-shot & 2-way 5-shot& 5-way 1-shot & 5-way 5-shot \\
\cmidrule(lr){2-5}

%\multirow{2}{c}{\textbf{Model}} &\multicolumn{4}{c}{VUW dataset} \\

%\multirow{2}{*}{Model}&\multicolumn{2}{c}{Population}&\multirow{2}{*}{Change (\%)}\\\cline{2-3}
%&2016&2017&\\

\multirow{2}{*}{Model}& 2-way 1-shot & 2-way 5-shot& 5-way 1-shot & 5-way 5-shot\\\cline{2-5}
&\multicolumn{4}{c}{VUW dataset}\\

%\multirow{2}{*}{\textbf{Model}}&\multicolumn{2}{c}{VUW dataset}&\\\\\

\hline
Relation  \cite{sung2018learning}  & 55.8 $\pm$  1.8\%& 55.2 $\pm$ 2.6\%   & 42.6 $\pm$ 3.1\%& 43.2 $\pm$ 2.1\%     \\
Prototypical  \cite{snell2017prototypical}   & 81.1 $\pm$ 2.2\%  & 86.4$\pm$ 1.9\% & \textbf{69.3} $\pm$ 2.2\% & \textbf{70.2} $\pm$ 2.0\%\\
Prototypical   \cite{snell2017prototypical}  (Gray)  & 74.2 $\pm$ 2.6\%  & 82.6 $\pm$ 2.5\%& 51.3 $\pm$ 2.9\% & 53.7 $\pm$ 2.1\%\\
Our model (No augmentation) & 80.1 $\pm$ 2.9\% & 83.1 $\pm$ 2.2\%& 60.2 $\pm$ 3.4 \% & 64.3 $\pm$ 2.8\%\\
Our model & \textbf{83.1}  $\pm$ 3.1\% & \textbf{87.2} $\pm$ 2.3\%& 63.2 $\pm$ 2.8\% & 68.1 $\pm$ 3.1\%\\
\hline
Prototypical (Obfuscation) &  80.3 $\pm$ 2.7\% & 84.8 $\pm$ 3.1\% & \textbf{65.2} $\pm$ 2.7\% & \textbf{68.3} $\pm$ 2.0\%\\

Our model (Obfuscation) & \textbf{81.7} $\pm$ 2.9\%\ & \textbf{85.2} $\pm$ 2.5\%\ & 53.2 $\pm$ 2.7\% & 57.3 $\pm$ 2.1\% \\

\midrule[0.3pt]
 & \multicolumn{4}{c}{Malimage dataset} \\
\hline
Relation \cite{sung2018learning}  & 63.2 $\pm$ 2.8\% & 68.1 $\pm$ 3.1\%& 56.8 $\pm$ 3.4\% & 58.2 $\pm$ 3.1\%\\
Prototypical \cite{snell2017prototypical} & 95.6 $\pm$ 1.9\% &  96.7 $\pm$ 1.8\%& \textbf{92.3} $\pm$ 2.3\% & \textbf{96.9} $\pm$ 2.1\%\\
Prototypical  \cite{snell2017prototypical} (Gray) & 94.3 $\pm$ 2.1\% &  94.9 $\pm$ 2.3\%& 90.1 $\pm$ 2.5\% & 94.8 $\pm$ 1.9\%\\
Our model (No augmentation) & 93.2 $\pm$ 1.8\%  & 95.2 $\pm$ 1.7\% & 80.1 $\pm$ 2.8\% & 87.1 $\pm$ 1.7\%   \\
Our model & \textbf{96.1} $\pm$ 0.7\%&  \textbf{97.3} $\pm$ 1.4\%& 83.8 $\pm$ 2.1\% & 90.2 $\pm$ 1.8\%\\
\hline
Prototypical (Obfuscation) &  91.9 $\pm$ 1.7\% & 94.7 $\pm$ 1.8\% & \textbf{85.4} $\pm$ 1.9\% & 85.7 $\pm$ 1.8\%\\

Our model (Obfuscation) & \textbf{94.3} $\pm$ 1.2\% &  \textbf{95.7} $\pm$ 1.1\%& 82.9 $\pm$ 2.2\% & \textbf{90.3} $\pm$ 2.1\%\\
\midrule[0.3pt]\bottomrule[1pt]
\label{results}
\end{tabular}

\label{table:results_init_trained}
\end{table*}

\subsection{Dataset setup and augmentation}
The number of malware samples from different malware classes in Table \ref{table:malware-family} vary significantly. Almost half of the classes had no more than 25 malware samples, while some only had one, likely because they were new malware samples detected recently (e.g., Blocal and Newbak). We increased the image sample size to allow for at least 30 samples for every malware class using a data augmentation technique (e.g., applying random transformations such as image rotations and re-scaling), in which the rotations are set up at degrees of 90, 180, and 270. At the same time the mean value and standard deviation were set to 0.52206 and 0.08426, as shown in Table \ref{table:parameters}.

We performed NOP insertion on the original image at frequencies 200 times each. Observed from Fig \ref{fig:entropy}, the same ransomware families have shown a similar entropy pattern.  Fig. \ref{fig:entropy} shows the visual difference in the obfuscated malware images, which also intuitively differs from the original entropy image. From this change, we can conclude that static signatures can be easily tampered by using a simple obfuscation technique such as NOP. However, the appearance of entropy images at frequencies with 200, 400, 600 does not show obvious visual differences between them. We conducted experiments on the 200-frequency dataset to evaluate the performance of our model.

We measured 1-shot and 5-shot accuracy in 2-way and 5-way on random datasets drawn from the total set of training and test sets in each of 20,000 episodes. The malware image size is fixed for each model.

\subsection{Training Details}
We adopt an episode-based training strategy, which takes the support set and query set into account during each training episode. This could be noted as $C$-way $K$-shot training. For example, the 5-way 1-shot contains one labeled sample for each unique class in 5-way. For $K$-shot where $K$ $>$ 1, we calculate the element-wise average over the embedding module outputs of all samples from each training class to form this class’ feature map. The class-level average pooling is then concatenated with the query image feature. The number of query images is dependent on the batch size in each training episode. For example, 2-way 5-shot contains 19 query images, the 5-way 5-shot has 15 query images. The corresponding relationship between the support samples, query samples, and batch size can be expressed by the following equation,
\begin{equation}
\begin{aligned}
S =& \{s^{(1)},..., s^{(c)}, ..., s^{(C)} \} \subset \mathbb{C}^{train}, s^{(c) } =K\\
Q = & \{q^{(1)},..., q^{(c)}, ..., q^{(C)} \} \subset \mathbb{B}^{train}, q^{(c) } = N
\ \\
\end{aligned}
\end{equation}
where $c$  is the class index and $K$ is the number of samples in class $s^{(c)}$; thus the training samples are constructed with $N \times K$ samples in each episode. Specifically, when we expand them, we can get the relationship below as:
\begin{equation}
\begin{aligned}
 \\ r_{ij}  : &\{(s^{(1)}, q^{(1)}),(s^{(1)}, q^{(1)}),(s^{(n)}, q^{(1)})\}, ... \ \  \\
\    & \{(s^{(1)}, q^{(c)}),(s^{(c)}, q^{(c)}),(s^{(n)}, q^{(c)})\} , ... \ \  \\
\   & \{(s^{(1)}, q^{(n)}),(s^{(c)}, q^{(n)}),(s^{(n)}, q^{(n)})\}  \ \  \\ 
\end{aligned}
\end{equation}
observed from this, the $(s^{(1)}, q^{(1)})$, $(s^{(c)}, q^{(c)})$, $(s^{(n)}, q^{(n)})$ are labelled as 1, others are labelled as 0.

The training sets and testing set are randomly split into the 9/6 classes and 2/5 classes respectively. The Malimages dataset is split into 13 classes for training and 12 classes for testing. We resized the entropy images to 105 $\times$ 105 and the results were gathered randomly 10 times. In one set the whole model ran 20000 epochs in the training stage and then was tested over 2000 epochs. The batch size of 19 and 15 were for the 1-shot and 5-shot respectively. Our model shared the initialized learning rate with the value of 0.02. Meanwhile, the hyperparameter for $\lambda$ in the loss function was 0.7. 

\subsection{Experimental Results on VUW Dataset}
\begin{figure*}
    \centering
    	\subfloat[Initialization Stage]
    	{\includegraphics[width=0.40\textwidth]{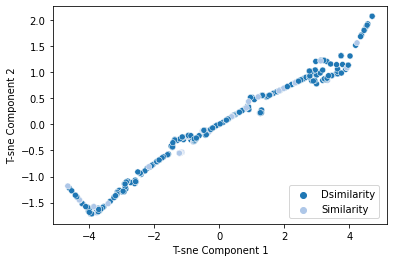}}\hfil
             \subfloat[Testing Stage]
    	{\includegraphics[width=0.40\textwidth]{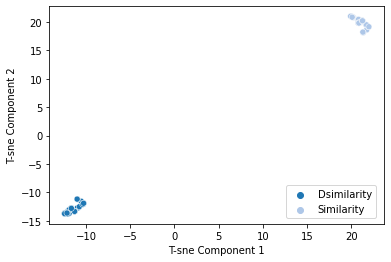}}\hfil  

        \caption{t-distributed stochastic neighbor embedding (t-SNE)  visualisation of embeddings generated using our model}
\label{t-sne-model}
\end{figure*}

As seen in Table \ref{table:results_init_trained}, we evaluated our model on two different features, the grayscale feature, and the entropy image. Our model achieved the highest accuracy with  83.1\% and 81.7\% on 2-way 1-shot and 2-way 5-shot, respectively. As a comparison, the prototype also has greater improvement with the entropy image on the 2-way and 5-way result than with the grayscale feature, which result increased by 7.1\%. The reason that our model and prototype had better performance than other models is that the representation of entropy images which exhibits higher uniqueness and entropy values incorporated into local regions with higher information content. It is worth noting that instead of taking the sum pooling for the feature concatenation like a relation network \cite{sung2018learning} does, we took the mean pooling for the few-shot learning. We further compared our performance with the prototype network under the obfuscated entropy image, and our model still showed a significant success on the 2-way results, which at 81.7\% and 85.2\% are higher than the prototype network by 1.4\% and 0.4\%, respectively. % However, the result on the 5-way drops significantly due to the large intra-class variance. Therefore, it is more difficult to construct a feature representation distinguished with a high inter-class variance, which also resulted in a greatly decreased 5-way result.
\begin{figure*}
    \centering
    	\subfloat[Average pooling]
    	{\includegraphics[width=0.40\textwidth]{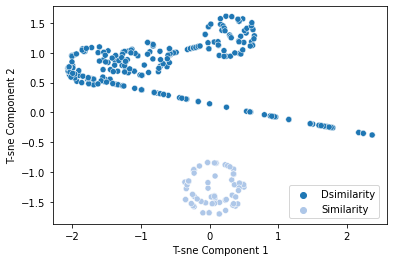}}\hfil
             \subfloat[Sum pooling]
    	{\includegraphics[width=0.40\textwidth]{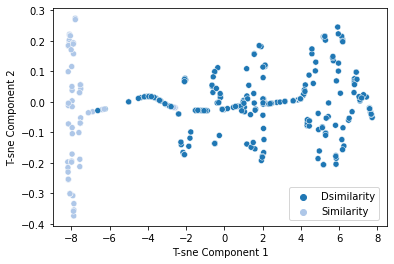}}\hfil  

        \caption{t-distributed stochastic neighbor embedding (t-SNE)  visualization of embeddings generated using our model}
\label{t-sne-model:sum_average}
\end{figure*}

\subsection{Experimental Results On Malimage Dataset}
It is worth noting that the intra-class variance across all samples within the Malimage dataset significantly decreased when compared with the VUW dataset. The feature representation can give better performance on the condition that there is a low level of intra-class variance. Our model still achieves the best performance on the 2-way result with 96.1\% and 97.3\% on 2-way 1-shot and 2-way 5-shot respectively on the non-obfuscated dataset. The performance on the obfuscated dataset is almost identical to the non-obfuscated dataset, which results in a drop of only 1.8\% and 1.6\% on 2-way 1-shot and 2-way 5-shot respectively. This is contrasted with the prototype network, which decreased 3.7\% and 2.0\%. The same situation happens on the 5-way performance, which resulted in the obfuscated dataset offering a drop of only 0.9\% on 5-way 1-shot. The performance on 5-way 5-shot only has a slight difference in result, whereas the prototype network decreased by 6.9\% and 11.2\% on the 5-way 1 shot and 5-way 5-shot on the obfuscated dataset. From observation of subtle differences in the results, we can prove that our model has a positive effect on the anti-obfuscation technology.

\subsection{Ablation Study}
\subsubsection{Hyperparameter $\lambda$ with the dimension of the linear layer} We first analyze how the performance  of our model is affected by the hyperparameter $\lambda$ and then we combine the dimensions of the linear layer in order to observe the changes in the results, as shown in Table \ref{table:nonlin-ablation}. It has been seen that the few-shot detection results are sensitive to the dimensions of the linear layer and the hyperparameter $\lambda$. A 256-dimension of linear layer and $\lambda$ = 0.7 achieve  better performance than other values, compared with the $\lambda$ = 0.3. %Secondly, our average feature calculated from  feature concatenation guides the relation-aware to construct an objective embedding that distinguishes the characteristics of intra-class and inter-class and reduces the rate of misclassifications. 
We analyzed the trend of different dimension combinations in the linear layer and the value of $\lambda$ for the performance. It can be seen from the Table \ref{table:nonlin-ablation}, that as the value of $\lambda$ increases while the dimension of the linear layer fixed, the accuracy rate increases gradually until the value of $\lambda$ reaches 0.9. When the value of $\lambda$ is fixed, the dimension of 256 performs better than with the dimension of 128.

\begin{table}[ht]
 % title of Table
\centering % used for centering table
\caption{The performence on Malimage dataset with hyperparameter $\lambda$ for  2-way 5-shot accuracy }
\begin{tabular}{c c c c c c c } % centered columns (4 columns)
\hline\hline %inserts double horizontal lines
 &  $\lambda = 0.3$ & $\lambda = 0.5$ & $\lambda = 0.7$ &  $\lambda = 0.9$ \\ [0.5ex] % inserts table 
  linear layer  &   &  \\ [0.5ex]
%heading
\hline % inserts single horizontal line
128 & 80.2 &  81.7  & 91.1 & 85.0\\ % inserting body of the table
256 &  82.7 & 86.4  & 95.7 & 86.1\\
\hline %inserts single line
\end{tabular}

\label{table:nonlin-ablation} % is used to refer this table in the text
\end{table}
\subsubsection{Pooling methods for feature selection}
We present a t-SNE visualization of the feature embeddings generated using our model. Based on different pooling methods used, Fig. \ref{t-sne-model} (a) illustrates the overall data points added with NOP obfuscation in the support set and query set, as projected into the raw embedding in which data points are optimized by episode training. The relation between the pair of samples is recognized from Fig. \ref{t-sne-model} (b). The two clusters are represented as similarity and dissimilarity, respectively, which offer the inter- and intra-characteristics. The model can benefit from the average calculation in the feature embedding since these averaging calculations are less deviated from the center of the object, as shown in Fig .\ref{t-sne-model:sum_average}. This figure clearly demonstrates that the data points in sum pooling are more dispersed, while the data points in average pooling are more structured.

\begin{figure}
    \centering
     {\includegraphics[width=0.40\textwidth]{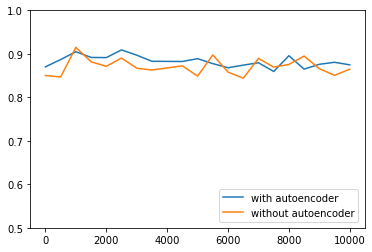}} \hfil

\caption{The comparison with or without autoencoder}
\label{albation:autoencoder}
\end{figure}

\subsubsection{Impact of autoencoder}
To explore the presence of the autoencoder in the schema, we use only part of the encoder in our model to learn the embedding representation while removing the decoder part. The encoder's purpose is to reduce the influence of the reconstruction loss function, as this is designed to create a robust feature against the NOP obfuscation. Observed from Fig \ref{albation:autoencoder}, the fluctuation range of the results obtained by the model without decoding is slightly larger than that with decoding while the average performance is decreased as well. The results without the decoder part are below those with the decoder part during the episodes ranging from 2000 to 6000. Moreover, its average result throughout the training phase is also slightly higher than that of the model without an autoencoder.

\section{Conclusion}

In this study, we introduce a new approach for detecting zero-day malware attacks that have been disguised using obfuscation techniques like junk code and no-operation code insertions. Our proposed solution, a Siamese Neural Network (SNN) combined with denoising autoencoders, addresses the challenge of identifying unseen malware signatures.

We take into account the relationships between features in different malware samples during the training process, leveraging entropy-based features to better capture the unique and structural information of each malware sample, even in the presence of obfuscation. Instead of relying on traditional distance scores, we use a relation-aware embedding method based on probabilities to accurately capture the semantic differences between malware samples and classify them.

Evaluations were conducted on two widely-used malware datasets, Malimage and VUW, to demonstrate the effectiveness of our proposed model in predicting unseen malware classes, even in the presence of obfuscation techniques.

In the future, we aim to broaden the scope of our research by incorporating a more extensive range of malware samples. This will allow us to uncover a more diverse range of correlations and further evaluate the adaptability and versatility of our proposed model. Additionally, we plan to incorporate various other obfuscation techniques such as code hiding using Packers/XOR/Base64, Register reassignment, and Code Transposition/Subroutine Reordering, among others. This will further enhance our model's ability to predict zero-day malware attacks effectively.


\section*{Acknowledgment}
This research is supported by the Cyber Security Research Programme—Artificial Intelligence for Automating Response to Threats from the Ministry of Business, Innovation, and Employment (MBIE) of New Zealand as a part of the Catalyst Strategy Funds under the
grant number MAUX1912.



%\bibliographystyle{IEEEtran}

\bibliography{references.bib}
\bibliographystyle{IEEEtran}

\end{document}